\documentclass[floatfix,a4paper]{article}
\pdfoutput=1
\usepackage{amsmath,amssymb}
\usepackage{dsfont}
\usepackage{graphicx} 
\usepackage{epstopdf} 
\usepackage{color}

\usepackage{graphicx,longtable,tocloft,color}
\usepackage{sidecap}
\usepackage{subfig}
\usepackage{xspace}
\usepackage{mydefs}
\usepackage{cite}
\usepackage{lineno}
\usepackage{multirow}
\usepackage{hyperref}
\usepackage[left=2.8cm,right=2.8cm,top=2.5cm,bottom=3.5cm]{geometry}
\usepackage{caption}
\usepackage{array}

\title{Constraining new physics with collider measurements of Standard Model signatures}

\author{Jonathan M. Butterworth$^1$, David Grellscheid$^2$,\\[1mm] Michael Kr\"amer$^3$, Bj\"orn Sarrazin$^3$, David Yallup$^1$\\[2.5mm]
\it $^1$Department of Physics and Astronomy, University College London,\\ \it Gower St., London, WC1E 6BT, UK\\[1mm]
\it $^2$IPPP, Department of Physics,\\\it Durham University, DH1 3LE, UK\\[1mm] \it $^3$Institute for Theoretical Particle Physics and Cosmology, \\ \it RWTH Aachen University, 52056 Aachen, Germany}

\begin{document}

\maketitle

\begin{abstract}
A new method providing general consistency constraints for Beyond-the-Standard-Model (BSM) theories, using
measurements at particle colliders, is presented. The method, `Constraints On New Theories Using \rivet', \Contur, exploits
the fact that particle-level differential measurements made in fiducial regions of phase-space have a high degree of
model-independence. These measurements can therefore be compared to BSM physics implemented in Monte Carlo generators in a very
generic way, allowing a wider array of final states to be considered than is typically the case. The \Contur approach should be seen
as complementary to the discovery potential of direct searches, being designed to eliminate inconsistent
BSM proposals in a context where many (but perhaps not all) measurements are consistent with the Standard Model.
We demonstrate, using a competitive simplified dark matter model, the power of this approach.
The \Contur method is highly scaleable to other models and future measurements.
\begin{flushright}IPPP-16-52, MCNET-16-21, TTK-16-22\end{flushright}
\end{abstract}

\section{Introduction}
\label{sec:intro}
The Large Hadron Collider (LHC) is probing physics in a new kinematic region, at energies around and above the
electroweak symmetry-breaking scale. With the discovery of the Higgs boson~\cite{Aad:2012tfa,Chatrchyan:2012ufa},
the first data-taking period of the LHC experiments demonstrated that the understanding of electroweak symmetry-breaking within
the Standard Model (SM) is broadly correct, and thus that the theory is potentially valid well above the
TeV scale. Many precision measurements of jets, charged leptons, and other final states
have been published, reaching into this new kinematic domain. The predictions of the SM are
generally in agreement with the data, while the many dedicated searches for physics beyond the SM
have excluded a wide range of possible scenarios.
Nevertheless, there are many reasons to be confident that
physics beyond the Standard Model (BSM) exists; examples include the gravitational evidence for dark matter, the large
preponderance of matter over antimatter in the universe, and the existence of gravity itself. None of
these can be easily accommodated within known Standard Model phenomenology.

This motivates a continued campaign to make precise measurements and calculations at higher energies and
luminosities, and to exploit these measurements to narrow down the class of viable models of new physics,
hopefully shedding light on the correct new theory, or at least on the energy scale at which
new physics might be observed at future experiments. Whether physics beyond the Standard Model is discovered
or not, there is a need to extract the clearest and most generic information about physics in this new energy regime,
an imperative which will grow with integrated luminosity.

In this paper we exploit three important developments to survey existing measurements and set
limits on new physics.

\begin{enumerate}
\item
SM predictions for differential and exclusive, or semi-exclusive, final states are made using sophisticated
calculational software, often embedded in Monte Carlo generators capable of simulating full, realistic final
states~\cite{Buckley:2011ms}. These generators now incorporate matrix-elements for higher-order processes
matched to logarithmic parton showers, and successful models of soft physics such as hadronisation and
the underlying event. They are also capable of importing new physics models into this framework, thus allowing
the rapid prediction of their impact on a wide variety of final states simultaneously.
In this paper we make extensive use of these capabilities within Herwig~7~\cite{Bellm:2015jjp,Bahr:2008pv}.
\item
As the search for many of the favoured BSM scenarios has been unsuccessful, there has been a move toward
``simplified models'' of new physics~\cite{Alves:2011wf,Abercrombie:2015wmb}, which aim to be as generic as possible
and which provide a framework for interpreting BSM signatures with a minimal amount of
new particles, interactions and model assumptions. The philosophy is similar to an ``effective lagrangian'' approach in which effective anomalous
couplings are introduced to describe new physics, but is more powerful, as such simplified models also
include new particles, and thus can remain useful up to and beyond the scale of new physics --- a region
potentially probed by LHC measurements.
\item
The precision measurements from the LHC have mostly been made in a manner which minimises their model-dependence.
That is, they are defined in terms of final-state signatures in fiducial regions well-matched to the
acceptance of the detector. Many such measurements are readily available for analysis and
comparison in the \rivet library~\cite{Buckley:2010ar}.
\end{enumerate}

These three developments together make it possible to efficiently
bring the power of a very wide range of data to bear on the search for new physics. While such a generic approach is
unlikely to compete in terms of speed and sensitivity with a search optimised for a specific theory, the breadth
of potential signatures and models which can be covered makes it a powerful complementary approach.\footnote{Limits from existing
searches can sometimes be applied to new models, for example by accessing archived versions of the original analysis code
and detector simulation via the RECAST~\cite{Cranmer:2010hk} project, or by independent implementations of experimental
searches, see, for example, Refs.~\cite{Conte:2012fm,Drees:2013wra,Kraml:2013mwa,Papucci:2014rja,Barducci:2014ila}.} On the one hand, any theory seeking
to explain a new signature or anomaly in the data may predict a BSM signal in other final states, which should be checked
against data this way. On the other hand, if no BSM physics emerges, a model-independent and systematic approach becomes mandatory to exclude
new physics models or narrow down the corresponding model parameter space.

In this paper, we first motivate and describe the
simplified model we have chosen as an initial demonstration, and the tools we use for its simulation.
In Section~\ref{sec:measurements} we introduce the measurements that we will use, and their implementation in \rivet.
Section~\ref{sec:method} covers the core of the \Contur method, incuding the statistical approach and dynamic data selection
and the assumptions made in this initial study.
In Section~\ref{sec:kinematics} we discuss the differential cross sections in which the impact of our example model would be
most apparent. In Section~\ref{sec:limits} this impact is translated into limits on the model parameters, and this is followed by our conclusions.

\section{Simplified Model}\label{sec:model}
Searches for new physics at the LHC are often interpreted in terms of simplified models. Simplified models provide a generic
framework for analysing experimental signatures using a small number of parameters, such as masses and
couplings of new fields, without reference to specific UV-complete models. Such an  approach is particularly
well-suited for interpreting the search for dark matter in a more model-independent way, and can be used to connect results from the LHC with
dark matter searches in direct detection and from the observation of cosmic rays. Many simplified models for dark matter have been proposed
in the past (see Ref.~\cite{Abercrombie:2015wmb} and references therein). Here, we consider a simplified model with a dark matter Majorana
fermion, $\psi$, which interacts with the SM model through a new vector particle, $Z^\prime$. The couplings of the mediator $Z^\prime$ to
the dark matter $\psi$ and to the
SM are specified as
\begin{align}\label{eq:vector_mediator}
 {\cal L} \supset  \GDM\,\overline{\psi} \gamma_{\mu}\gamma_5 \psi\,Z'^{\mu} + \GQ\sum_{q} \bar q \gamma_{\mu} q \,Z'^{\mu} \,,
\end{align}
where the sum in the second term includes the first generation SM quarks, $q \in \{u,d\}$. The simplified model specified in Eq.~(\ref{eq:vector_mediator})
has only four free parameters, two couplings and two masses: \GDM, \GQ, $M_\psi \equiv \MDM$, and \MZP. The width of the mediator,
$\Gamma_{Z'}$, is determined by these four parameters.

Following Ref.~\cite{Kahlhoefer:2015bea} we have chosen to couple the mediator to
dark matter and to the SM quarks through an axial-vector and vector current, respectively. An axial-vector coupling of the
mediator to dark matter leads to spin-dependent dark matter-nucleon interactions and thus weaker bounds from direct dark matter searches.
Such a coupling structure naturally arises for Majorana fermion dark matter. Having also axial-vector couplings between
the mediator and the SM  requires UV-completions of the simplified model in which the SM Higgs has to be charged under
the $U(1)'$ gauge group of the vector mediator. As a consequence, there is mixing between the $Z'$ and the gauge boson of the SM, and gauge invariance
requires the couplings of the $Z'$ to be flavour universal. However, models where the mediator couples to leptons
are strongly constrained by collider searches for di-lepton resonances. For vector couplings of the $Z'$ to  SM fermions, on the other hand,
the SM Higgs does not carry a $U(1)'$ charge and the charges of quarks and leptons are independent. We are thus free to set the
$Z'$ lepton coupling to zero to evade constraints from di-lepton searches, and consider a simplified model with a universal vector-coupling
to SM quarks only.

The parameters of the simplified model, \GDM, \GQ, \MDM, and \MZP, are constrained by perturbative unitarity.
From a partial wave analysis of the annihilation process $\overline{\psi}\psi \to Z' Z'$ one can derive the unitarity limit~\cite{Kahlhoefer:2015bea}
\begin{equation}
\MDM \lesssim \sqrt{\frac{\pi}{2}} \frac{\MZP}{g_{\rm DM}}\,,
\end{equation}
which defines the parameter space where the dark matter relic density can be reliably calculated within the simplified model.
Perturbative unitarity of the scattering amplitude in processes relevant to LHC dark matter searches has been studied in Ref.~\cite{Englert:2016joy}.
It was found that perturbative unitarity is respected in the production of mediators at the LHC, unless
the couplings are large, $\GQ \gtrsim {\cal O}(4\pi)$. In our analysis, we will only consider couplings which are well within the
perturbative regime, \GDM, $\GQ \lesssim {\cal O}(1)$, so that our predictions for dark matter and mediator production at the LHC
are well-defined.

Dark matter has been searched for at the LHC in signatures with jets and large missing transverse momentum, see e.g.~\cite{Aaboud:2016tnv,CMS:2016tns}
for recent analyses.
The results~\cite{Aaboud:2016tnv,CMS:2016tns}  have not been interpreted in the simplified model defined in Eq.~(\ref{eq:vector_mediator}), but in
similar models with pure vector or axial-vector mediators and Dirac fermion dark matter. The  searches probe the region where $\MDM \lesssim \MZP/2$
and exclude dark matter and mediator masses of up to about 500~GeV and 1.5~TeV, respectively. Similar exclusions have been obtained in simplified
model re-interpretations of LHC searches as presented in, e.g.,  Refs.~\cite{Kahlhoefer:2015bea, Heisig:2015ira}. Searches for dijet resonances from
mediator production and decay can place further strong constraints on the dark matter simplified model as demonstrated in Refs.~\cite{Chala:2015ama,Fairbairn:2016iuf}.

To simulate the experimental signature for our model, we have encoded the model Lagrangian in \textsc{FeynRules}~2.3.18~\cite{Alloul:2013bka}. Using its UFO interface~\cite{Degrande:2011ua},
a BSM configuration is created for Herwig~7.0.1~\cite{Bellm:2015jjp,Bahr:2008pv}. For each parameter point in the scan grid, events were generated in Herwig and
analysed using the selected analyses implemented in
 \rivet 2.4.1~\cite{Buckley:2010ar} (see section \ref{sec:measurements}). Calculation of the exclusion contours was done in Python scripts, available on the \Contur website \url{https://contur.hepforge.org/}.

\begin{figure}[hp]
\centering
\subfloat[\label{fig:schannel}]{
\includegraphics[width=0.3\textwidth]{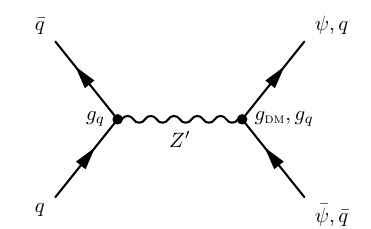}
}
\subfloat[\label{fig:plusjet}]{
\includegraphics[width=0.3\textwidth]{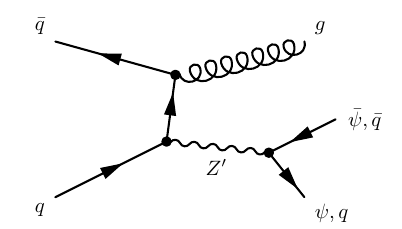}
}
\subfloat[\label{fig:plusV}]{
\includegraphics[width=0.3\textwidth]{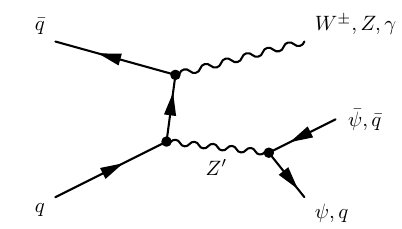}
}
\caption{Relevant Feynman diagrams introduced by the simplified model at leading order. (a) $s$-channel production followed by decay to quarks or to DM, (b) associated jet production (c) associated gauge-boson production.}\label{fig:feyndia}
\end{figure}

Higher-order QCD corrections have been calculated for this class of dark matter simplified models
with an $s$-channel vector mediator, see Refs.~\cite{Fox:2012ru,Haisch:2013ata,Backovic:2015soa,Neubert:2015fka}. However, for the purpose of this paper where we focus on introducing the \Contur approach rather than exploring
a particular BSM theory in great detail, we will use leading-order signal cross section predictions as provided by Herwig 7.
The most relevant production and decay channels for the mediator are illustrated in Fig.~\ref{fig:feyndia}.

\section{Measurements}\label{sec:measurements}

To be useful in our approach, measurements must be made in as model-independent a fashion as possible.
Cross sections should be measured in a kinematic region closely matching the detector acceptance --- commonly called
`fiducial cross sections' --- to avoid extrapolation into unmeasured regions, since such extrapolations must always make
theoretical assumptions; usually that the SM is valid. The measurements should generally be made in terms of observable final
state particles (e.g. leptons, photons) or objects constructed from such particles (e.g. hadronic jets, missing energy)
rather than assumed intermediate states ($W, Z, H$, top). Finally, differential measurements are most useful, as features
in the shapes of distributions are a more sensitive test than simple event rates --- especially when there are
highly-correlated systematic experimental uncertainties, such as those on the integrated luminosity, or the jet energy scale.

One feature noted in several cases is that missing transverse energy (\MET) is
explicitly assumed to be the same as neutrino transverse energy. In BSM physics, missing energy can also
arise from other sources (for example, dark matter production) and so it is important that the result is treated in such a
way that this sensitivity is correctly estimated. The measurements are typically corrected back to total \MET,
or to the assumed neutrino \pt, in the experimental analysis, using a simulated SM event sample which has been shown to describe
the data well. This involves an extrapolation into the forward
region where transverse energy is unmeasured; however, unless a BSM particle enters this region, the error made is
negligible. This means that as long as (in the \rivet analysis) a fiducial acceptance cut is made on BSM particles
counting toward \MET (to ensure that large contributions to \MET
from invisible particles outside the detector acceptance are excluded) such measurements can be
used.\footnote{Of greater consequence, but easier to fix, is the fact that several \rivet methods explicitly calculated \MET
from neutrinos found in the simulated event record, rather than as the negative of the visible particles in the event. These
routines were modified as a part of this work, and are fixed in future \rivet releases.}

Another feature of the measurements is that most of them, explicitly or implicitly, insist in their fiducial cross-section definition
that leptons and photons be `directly' produced, that is, prior to hadronisation and coming from the primary vertex of the collision.
Such a selection is enforced in the experiments by a mixture of isolation and vertex requirements, but is not universally enforced
in all \rivet routines. Generally this is a small effect, but care needs to be taken that the sensitivity is not overestimated, especially
for BSM models which enhance bottom or charm production, when semi-leptonic decays may play a role.
This feature will be addressed in future releases of \rivet.

The measurements we consider fall into five loose and independent classes.
\begin{enumerate}
\item
Jets: event topologies with any number of jets but no missing energy, leptons, or photons. In this category there
are important measurements from both ATLAS and CMS, many of which have existing \rivet analyses. We make use of
the highest integrated-luminosity inclusive~\cite{Aad:2014vwa,Chatrchyan:2014gia}, dijet~\cite{Aad:2013tea,Aad:2014pua}
and three-jet~\cite{Aad:2014rma}
measurements made in 7~TeV collisions, as well as the jet mass
measurement from CMS~\cite{Chatrchyan:2013vbb}.
Unfortunately results from 8~TeV collisions are rarer, and the only one we can use currently is the four-jet
measurement from ATLAS~\cite{Aad:2015nda}.
\item
Electroweak: events with leptons, with or without missing energy or photons. The high-statistics $W+$jet and $Z+$jet measurements
from ATLAS~\cite{Aad:2014qxa,Aad:2013ysa}
and CMS~\cite{Khachatryan:2014uva,Khachatryan:2014zya}, are used.
We also use the ATLAS $ZZ$ and $W/Z+\gamma$ analyses~\cite{Aad:2012awa,Aad:2013izg}, the former of which includes \MET,
via the $Z \rightarrow \nu\bar{\nu}$ measurement.
\item
Missing energy, possibly with jets but no leptons or photons. This channel could in principle provide powerful constraints, and has been
used in searches (see for example~\cite{Aad:2012fqa}). Unfortunately however, there are currently no fully-corrected particle-level
distributions available in this category.
\item
Isolated photons, with or without missing energy, but no leptons. Here we make use of the inclusive~\cite{Aad:2013zba}, diphoton~\cite{Aad:2012tba} and
photon-plus-jet~\cite{ATLAS:2012ar} measurements,
where available. We also made a new \rivet routine for the CMS photon-plus jet measurement \cite{Chatrchyan:2013mwa}.
\item
Signatures specifically based on top quark or Higgs candidates. Most such measurements to date have been made at the 'parton' level (that is,
corrected using SM MC back to the top or Higgs before decay), and many of them are extrapolated to $4\pi$ phase space. Both steps increase
the model dependence and make them unsuitable for the \Contur approach. Recently, however, fiducial, differential, particle-level measurements
have begun to appear\cite{Aad:2015hna,Khachatryan:2016gxp}. These are potentially very powerful in excluding some models, but will in principle
overlap with the previous
categories depending on decay mode. We leave the inclusion of such measurements for future work.
\end{enumerate}

The choice of which measurements are actually included at this stage is driven mainly by the availability of particle-level differential
fiducial cross sections
implemented in \rivet. The current selection is summarised in Table~\ref{tab:Rivet}.

\newcommand{\breaker}{}

\renewcommand{\arraystretch}{1.5}

\begin{table}[hp]
\footnotesize
\begin{center}
\begin{tabular}{l  l  p{8cm} }
 \Contur Category & \rivet / Inspire ID & \rivet description \\
\hline
\hline
ATLAS 7 Jets  & ATLAS\_2014\_I1325553 \cite{Aad:2014vwa} & Measurement of the inclusive jet cross-section \\
& ATLAS\_2014\_I1268975 \cite{Aad:2013tea} & High-mass dijet cross section \\
& ATLAS\_2014\_I1326641 \cite{Aad:2014rma} & 3-jet cross section \\
& ATLAS\_2014\_I1307243 \cite{Aad:2014pua} & Measurements of jet vetoes and azimuthal decorrelations in dijet events \\
\breaker
CMS 7 Jets  & CMS\_2014\_I1298810 \cite{Chatrchyan:2014gia} & Ratios of jet pT spectra, which relate to the ratios of inclusive, differential jet cross sections \\
\breaker
ATLAS 8 Jets  & ATLAS\_2015\_I1394679 \cite{Aad:2015nda}  & Multijets at 8 TeV \\
\breaker
ATLAS 7  $Z$ Jets & ATLAS\_2013\_I1230812 \cite{Aad:2014qxa} & $Z$ + jets \\
\breaker
CMS 7 $Z$ Jets & CMS\_2015\_I1310737 \cite{Khachatryan:2014zya} & Jet multiplicity and differential cross-sections of $Z$+jets events \\
\breaker
CMS 7 $W$ Jets  & CMS\_2014\_I1303894 \cite{Khachatryan:2014uva} & Differential cross-section of $W$ bosons + jets \\
\breaker
ATLAS 7 $W$ jets & ATLAS\_2014\_I1319490 \cite{Aad:2013ysa} & $W$ + jets \\
\breaker
ATLAS 7 Photon Jet & ATLAS\_2013\_I1263495 \cite{Aad:2013zba} & Inclusive isolated prompt photon analysis with 2011 LHC data \\
 & ATLAS\_2012\_I1093738 \cite{ATLAS:2012ar} & Isolated prompt photon + jet cross-section \\
\breaker
CMS 7 Photon Jet  & CMS\_2014\_I1266056 \cite{Chatrchyan:2013mwa} & Photon + jets triple differential cross-section \\
\breaker
ATLAS 7 Diphoton & ATLAS\_2012\_I1199269 \cite{Aad:2012tba} & Inclusive diphoton $+ X$ events \\
\breaker
ATLAS 7 $ZZ$  & ATLAS\_2012\_I1203852 \cite{Aad:2012awa} & Measurement of the $ZZ(*)$ production cross-section \\
\breaker
ATLAS $W$/$Z$ gamma & ATLAS\_2013\_I1217863 \cite{Aad:2013izg}  & $W$/$Z$ gamma production \\
\hline
\hline
\end{tabular}
\caption{Table of all \rivet routines currently included in the limit-setting scan. With the one indicated exception, they are all based on 7~TeV data.}\label{tab:Rivet}
\end{center}
\normalsize
\end{table}
\renewcommand{\arraystretch}{1}

\section{Method}
\label{sec:method}

\subsection{Strategy}\label{sec:strat}

The approach taken is to consider simplified BSM models in the light of existing measurements which have {\it already been shown to agree} with SM expectations.
Thus this is inherently an exercise in limit-setting rather than discovery. The assumption is that a generic, measurement-based approach such as this will not be
competitive in terms of sensitivity, or speed of discovery, with a dedicated search for a specific BSM final-state signature. However, it will have the
advantage of breadth of coverage, and will make a valuable contribution to physics at the energy frontier whether or not new signatures are discovered at the LHC.
In the case of a new discovery, many models will be put forward to explain the data, as has for example already been seen\cite{PhysRevLett.116.150001}
after the 750~GeV diphoton anomaly
reported by ATLAS and CMS at the end of 2015 and start of 2016~\cite{ATLAS-CONF-2016-018,CMS-PAS-EXO-16-018}.
Checking these models for consistency with existing measurements will be vital for unravelling whatever
the data might be telling us. As will be shown in subsequent sections, models designed to explain one signature may have somewhat unexpected consequences
in different final states, some of which have already been precisely measured. If it should turn out that no BSM signatures are in the end confirmed at the LHC,
\Contur offers potentially the broadest and most generic constraints on new physics, and motivates the most precise possible model-independent measurements
over a wide range of final states, giving the best chance of an indirect pointer to the eventual scale of new physics.
Given this strategy, possible treatments of the data present themselves. The most complete is to take precision SM calculations to define the background,
with their associated uncertainties, and superimpose the putative signal, and check for consistency with the data within uncertainties. However, for striking signals
such as those considered here, and for data which have already been shown to exhibit no such striking features and indeed to agree with SM calculations,
it is reasonable and much more efficient to make the assumption for such measurements that the data are the SM, and to take the uncertainties on the data as defining the
room that is left for BSM signatures. Neither approach treats interference effects properly -- this would require a ful final-state calculation including all
SM and BSM diagrams, which are in general not available.
In this initial study, we follow the second approach, although future plans include incorporating SM predictions and their uncertainties directly.

\subsection{Dynamical data selection}\label{sec:dynselec}
\label{sec:selec}

Starting with the measurements discussed in Section~\ref{sec:measurements}
we define a procedure to combine exclusion limits from different measured distributions.
The data used for comparison in \rivet come in the form of histograms, which do not carry information about the correlations between uncertainties ---
even when in several cases detailed information is made available in the experimental papers. There are highly correlated uncertainties in several measurements,
for example on the integrated luminosity, or the energy scale of jet measurements. In some cases these are dominant. Including correlations would be a highly
complex process, since as well as correlations within a single data-set, there are also common systematic uncertainties between different
results, which are generally not provided by the experiments. There are also overlaps between event samples used in many different measurements, which lead to
non-trivial correlations in the statistical uncertainties. To attempt to avoid spuriously high exclusion rates due to multiply-counting what might be the
same exclusion against several datasets, we take the following approach:
\begin{enumerate}
\item Divide the measurements into groups that have no overlap in the event samples used, and hence no statistical correlation
between them. These measurements are grouped by, crudely, different final states, different experiments, and different beam
energies (see Table~\ref{tab:Rivet}).
\item Scan within each group for the most significant deviation between BSM+SM and SM. This is done distribution-by-distribution and bin-by-bin within distributions.
Use only the most significant deviation, and disregard the rest. Although the selection of the most significant deviation sounds intuitively suspect, in our case it
is a conservative approach, since we make the assumption that the data are equal to the SM, and discarding the less-significant bins simply reduces sensitivity.
The use of a single bin from each measured distribution removes the dominant effect of highly correlated
systematic uncertainties within a single measurement. Where several
of statistically-independent distributions exists {\it within} a group, their likelihoods may be combined to give a single likelihood ratio from the group, on the assumption that the systematic correlations between distributions are reduced compared to those within a single distribution.
\item Combine the likelihood ratios of the different groups to give a single exclusion limit.
\end{enumerate}

\subsection{Statistical Method}\label{sec:statmethod}
The question we wish to ask of any given BSM proposal is {\it `at what significance do existing measurements, which agree with the SM, already exclude this'}.
For all the measurements considered, comparisons to SM calculations have shown consistency between them and the data. Thus as a starting point, we take the
data as our ``null signal'', and we superpose onto them the contribution from the BSM scenario under consideration. The uncertainties on the data will define the
allowed space for these extra BSM contributions.

Taking each bin of each distribution considered as a separate statistic to be tested, a likelihood function for each bin can be constructed as follows,
\begin{align}\label{eq:likely}
L(\mu, {b}, {\sigma}_{b}, {s}) = { \frac{(\mu s + b)^{n}}{n!} \exp\big(-(\mu s + b)\big) \times \frac{1}{\sqrt{2 \pi} \sigma_{b}} \exp\left(-\frac{(m - b)^{2}}{2 \sigma_{b}^{2}}\right)} \times \frac{(\tau s)^{k}}{k!}\exp\big(-\tau s\big)\,,
\end{align}
where the three factors are:
\begin{itemize}
\item A Poisson event count, noting that the measurements considered are differential cross section measurements, hence the counts are multiplied by a factor of the integrated luminosity taken from the experimental paper behind each analysis, to convert to an event count in each bin (and subsequently the additional events that the new physics would have added to the measurement made). This statistic in each tested bin then is comprised of:
\begin{itemize}
\item $s$, the parameter defining the BSM signal event count. 
\item $b$, the parameter defining the background event count.
\item $n$, the observed event count.
\item $\mu$, the signal strength parameter modulating the strength of the signal hypothesis tested, thus $\mu=0$ corresponds to the background-only hypothesis
and $\mu=1$ the full signal strength hypothesis;
\end{itemize}
\item A convolution with a Gaussian defining the distribution of the background count, where the following additional components are identified:
\begin{itemize}
\item $m$, the background count. The expectation value of this count, which is used to construct the test, is taken as the central value of the measured data point.
\item $\sigma_{b}$, the uncertainty in the background event count taken, from the data, as 1 $\sigma$ error on a Gaussian (uncertainties taken as the combination of statistical and systematics uncertainties in quadrature. Typically the systematic uncertainty dominates).
\end{itemize}
\item An additional Poisson term describing the Monte Carlo error on the simulated BSM signal count with $k$ being the actual number of generated BSM events. The expectation value of $k$ is related to $s$ by a factor $\tau$, which is the ratio of the generated MC luminosity to the experimental luminosity.

\end{itemize}
This likelihood is then used to construct a  test statistic based on the profile likelihood ratio, following the arguments laid out in Ref.~\cite{Cowan:2010js}. In particular, the $\tilde{q}_{\mu}$ test statistic is constructed.
This enables the setting of a one-sided upper limit on the confidence in the strength parameter hypothesis, $\mu$,
desirable since in the situation that the observed strength parameter exceeds the tested hypothesis, agreement with
the hypothesis should not diminish.
In addition this construction places a lower limit on the strength parameter, where any observed fluctuations below the backgrund-only
hypothesis are said to agree with the background-only hypothesis~\footnote{At present, the latter point will be unimportant, as the manner in
which samples are generated and tested will only increase the event rates with respect to the background-only hypothesis.}.
The required information then is the sampling distribution of this test statistic. This can either be evaluated either using the so called Asimov data set to build an
approximate distribution of the considered test statistic, or explicitly using multiple Monte Carlo `toy model' tests~\footnote{For the cases considered here the results were found to be equivalent, implying that the tested parameter space values fall into the asymptotic, or large sample, limit, and so the Asimov approach is used.}.
\\

The information needed to build the approximate sampling distributions is contained in the covariance matrix composed of the second derivatives with respect to the parameters ($\mu, b$ and $s$), of the log of the likelihood given in equation \ref{eq:likely}. They are as follows:
\begin{align*}\label{eq:likely2}
  \mu \mu :& &\frac{\partial^2{\text{ln}L}}{\partial{\mu^2}} = & \frac{-ns^2}{(\mu s + b)^2}  \\
 b b :& &\frac{\partial^2{\text{ln}L}}{\partial{b^2}} = & \frac{-n}{(\mu s + b)^2} - \frac{1}{\sigma_b^2} \\
 s s :& &\frac{\partial^2{\text{ln}L}}{\partial{s^2}} = & \frac{-n\mu^2}{(\mu s + b)^2} - \frac{k}{s^2} \\
 \mu s = s \mu :& &\frac{\partial^2{\text{ln}L}}{\partial{\mu \partial s}} = & \frac{nb}{(\mu s + b)^2} - 1 \\
 \mu b = b \mu :& &\frac{\partial^2{\text{ln}L}}{\partial{\mu \partial b}} = &\frac{-ns}{(\mu s + b)^2} \\
 b s = sb :& &\frac{\partial^2{\text{ln}L}}{\partial{s \partial b}} =& \frac{-n\mu}{(\mu s + b)^2}.
\end{align*}
Which are arranged in the inverse covariance matrix as follows.
\begin{equation}
\label{eq:covar}
V^{-1} = - E
\begin{bmatrix}
    \mu\mu & \mu s & \mu b   \\
    s \mu & s s & s b    \\
    b \mu & b s &  b b

\end{bmatrix}
\end{equation}

The variance of $\mu$ is extracted from the inverse of the matrix given in eq.\ref{eq:covar} as;
\begin{equation}
\sigma_\mu^{2} = V_{\mu,\mu}
\end{equation}

In order to evaluate this, the counting parameters ($n, m$ and $k$) are evaluated at their Asimov values, following arguments detailed in Ref.~\cite{Cowan:2010js}
. These are taken as follows,
\begin{itemize}
\item $n_{A} = E[n] = \mu' s + b$. The total count under the assumed signal strength, $\mu'$, which for the purposes of this argument is equal to 1
\item  $m_{A}=E[m] = b$. The background count is defined as following a Gaussian distribution with a mean of $b$.
\item $k_{A} = E[k] = \tau s$. The signal count is defined following a Poisson distribution with a mean of $\tau s$
\end{itemize}
Using this data set the variance of the strength parameter, $\mu$, under the assumption of a hypothesised value, $\mu'$, can be found. This is then taken to define the distribution of the $\tilde{q}_{\mu}$ statistic, and consequently the size of test corresponding to the observed value of the count. The size of the test can be quoted as a $p$-value, or equivalently the confidence level which is the inverse of the size of the test.
As is convention in the particle physics community, the final measure of statistical agreement is presented in terms of what is known as the CL$_{s}$
method~\cite{Junk:1999kv,Read:2002hq}. 
Then, for a given distribution, CL$_{s}$
can be evaluated separately for each bin, where the bin with the largest CL$_{s}$ value (and correspondingly smallest $p_{s+b}$ value) is taken to represent the
sensitivity measure used to evaluate each distribution, a process outlined in section~\ref{sec:selec}.

Armed then with a list of selected sensitive
distributions with minimal correlations, a total combined CL$_{s}$ across all considered channels can then be constructed from
the product of the likelihoods. This leaves the core of the methodology presented here unchanged, the effect is simply extending the covariances matrix.
%
The overall result gives a probability, for each tested parameter set, that
the observed counts $n_{i}$, across all the  measurement bins considered, are compatible with the full signal strength hypothesis.

Finally it is noted that this methodology has been designed to simply profile BSM contributions against data taken. This can be extended to incorporate a separate background simulation or include correlation between bins where available.

\subsection{Limitations}

We note that our method is best adapted to identifying kinematic features (mass peaks, kinematic edges) and will be less sensitive to smooth
deviations in normalisation. In particular, since we take the data to be identically equal to the SM expectation, we will be insensitive
to a signal which might in principle arise as the cumulative effect of a number of statistically insignificant deviations
across a range of experimental measurements.
No such effects are apparent when studying the model considered here, but quantifying this statement is beyond the scope of the current work,
and requires an extensive evaluation of the theoretical uncertainties on the SM predictions for each channel. This is an extension
of the method planned for future work.
Additionally, in low statistics regions, outlying events in the tails of the data will not lead to a weakening of the limit,
as would be the case in a search. However, measurements unfolded to the particle-level are typically performed in bins with a
requirement of minimum number of events in any given bin, reducing the impact of this effect. Our limits focus on the impact
of high precision measurements on the BSM model, in which systematic uncertainties typically dominate. For these reasons, the limits derived are described as expected limits, although in regions where the confidence level is high, they do represent a real exclusion.

\section{Comparison to Data}\label{sec:kinematics}

To investigate the exclusion power of the SM measurements discussed in Section~\ref{sec:measurements} we scan a range in
plausible mediator masses (\MZP) and dark matter masses (\MDM) within the model described in Section~\ref{sec:model}, for three choices of
the coupling of the mediator to the SM (\GQ). These coupling choices correspond to (i) an `optimistic' scenario $\GQ = 0.5, \GDM = 1$: strong signals,
close to the edge of exclusion already, (ii) a `challenging' scenario $\GQ = 0.25, \GDM = 1$: low couplings, hard to exclude,
and (iii) an `intermediate' scenario $\GQ = 0.375, \GDM = 1$, between the two. We also consider (iv) a scenario where the coupling of dark matter to the mediator is
suppressed, $\GQ = 0.375, \GDM = 0.25$. For all these scenarios, the calculated width of the mediator is less than 10\% of \MZP, as shown in Table~\ref{tab:width}.

\renewcommand{\arraystretch}{1.5}
\begin{table}[hp]
\begin{center}
\begin{tabular}{c | c | c | c | c}
\GQ & \GDM & \MZP [GeV] & \MDM [GeV] & $\Gamma_{Z'}/\MZP$ \\
\hline
0.25 & 1 & 3000 & 100 & 0.0626 \\
0.375 & 1 & 3000 & 100 & 0.0751 \\ 
0.5 & 1 & 3000 & 100 & 0.0925 \\ 
0.375 & 0.25 & 3000 & 100 & 0.0257 \\ 
\end{tabular}
\caption{Table of maximal $\Gamma_{Z'}/\MZP$ occuring over the mass ranges for the four heatmaps shown in Figure~\ref{fig:maps}}\label{tab:width}
\end{center}
\end{table}
\renewcommand{\arraystretch}{1}

\begin{figure}[htb]
\centering
      \includegraphics[width=0.9\textwidth]{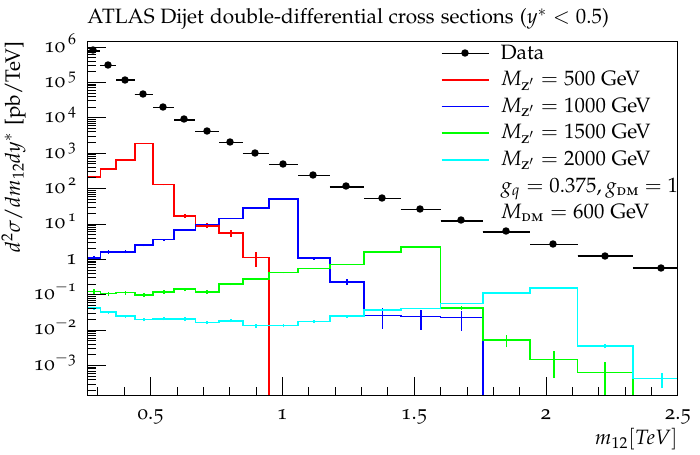}\\
      \includegraphics[width=0.9\textwidth]{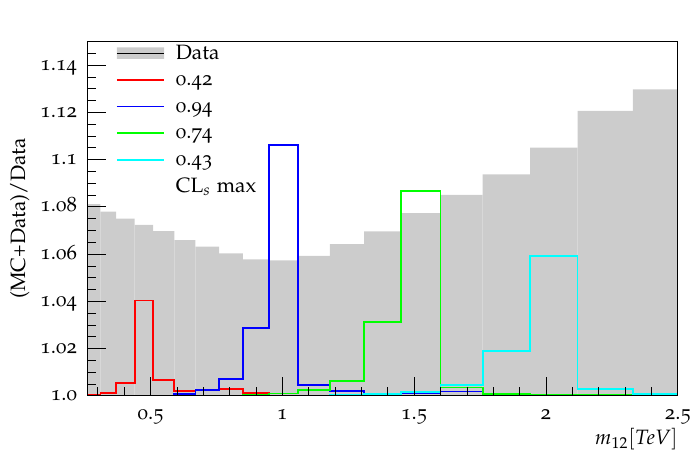}
    \caption{ Outputs from \rivet for a measurement included in the limit setting process. Simulated signals for a sample of mediator masses are shown, superimposed
on the double differential inclusive jet cross section in the most central rapidity region, binned by dijet mass and rapidity as measured by ATLAS at
7 TeV~\cite{Aad:2014pua}. The upper plot compares the measured cross section to the model expectation, and the lower hand plot shows the perturbation
in the ration compared to the relative uncertainty in the measurement.
The signals form a 1D parameter space scan in mediator mass for fixed dark matter mass and mediator couplings; $\MDM=600$ GeV,
$\GQ=0.375$ and $\GDM = 1$. The corresponding exclusion limits are also given.}
\label{fig:ATLASdijet}
\end{figure}

Figure~\ref{fig:ATLASdijet} shows a comparison between the model expectations in the `intermediate' scenario and the most sensitive distributions from the ATLAS jet measurements.
The measured dijet mass distribution is smoothly falling to higher masses, and the presence of a mediator decaying to quarks
(see Fig.\ref{fig:schannel} and \ref{fig:plusjet}) would superimpose a
peak, not seen in the data, thus leading to an exclusion. The results are shown for fixed $\MDM = 600$~GeV and a range of mediator massses $500 < \MZP < 2000$~GeV.
The sensitivity is at maximum in the middle of this range.

\begin{figure}
\centering
      \includegraphics[width=0.9\textwidth]{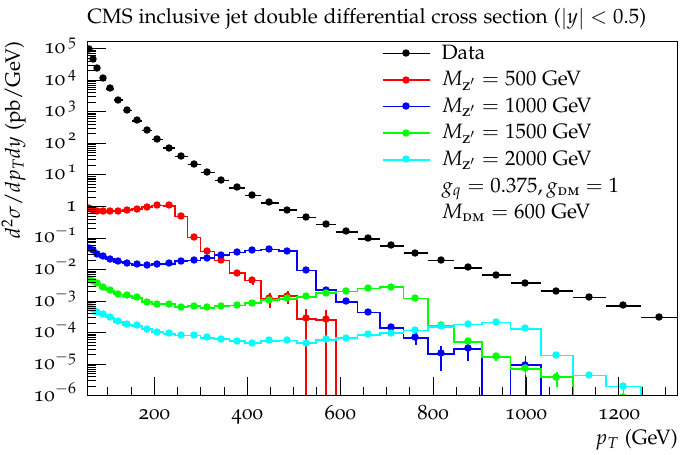}
      \includegraphics[width=0.9\textwidth]{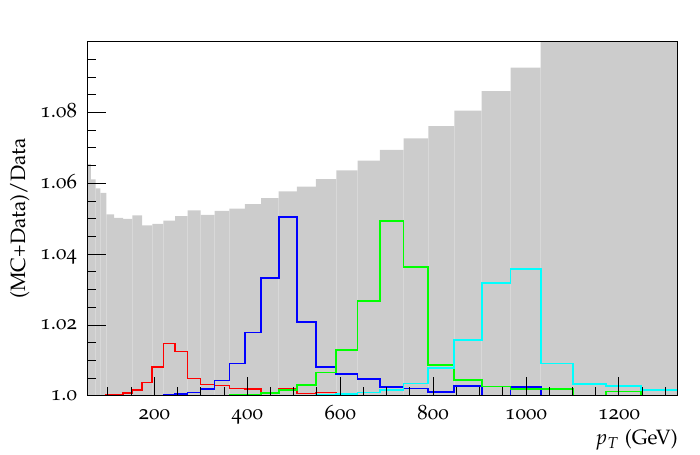}
\caption{Outputs from \rivet for a measurement included in the limit setting process. Simulated signals for a sample of mediator masses are shown, superimosed on
the double differential dijet cross section in the most central rapidity region, binned by leading jet \pt and rapidity as measured by
CMS at 7 TeV~\cite{Chatrchyan:2014gia}. The upper plot compares the measured cross section to the model expectation, and the lower plot shows the perturbation
in the ration compared to the relative uncertainty in the measurement. The signals form a 1D parameter space scan in mediator mass for fixed dark matter mass
and mediator couplings; $\MDM=600$ GeV, $\GQ=0.375$ and $\GDM = 1$.}
\label{fig:CMSincljet}
\end{figure}

Figure~\ref{fig:CMSincljet} shows a similar comparison for a comparable measurement from CMS. This time the sensitivity is in the jet \pt distribution, but the pattern
is similar, with a maximal sensitivity for mediator masses around 1~TeV. These measurements typify the sensitivies obtained from the `Jets' measurements discussed in
Section~\ref{sec:measurements}. It is notable that 7~TeV measurements form the bedrock of the exclusions. This is due to the lack of availability of precision
8~TeV and 13~TeV measurements in \rivet. Such measurements are likely to be available soon and can be expected to significantly improve the exclusion power of these
final states.

\begin{figure}
\centering
      \includegraphics[width=0.9\textwidth]{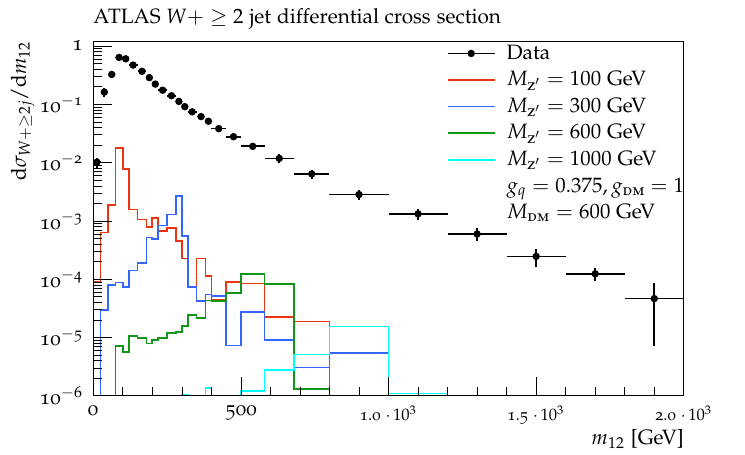}
      \includegraphics[width=0.9\textwidth]{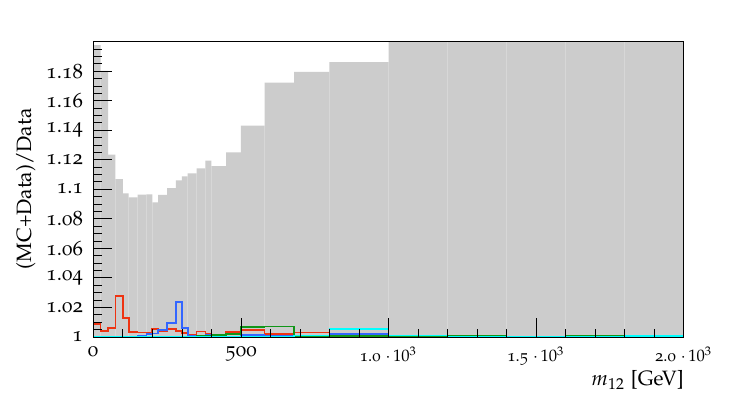}
\caption{Outputs from \rivet for a measurement included in the limit setting process. Simulated signals for a sample of mediator masses, superimposed on
the differential cross section for the $W+\geq2$ jet process, binned in the mass of the dijet pair as measured by ATLAS at 7 TeV~\cite{Aad:2014qxa}.
The signals form a 1D parameter space scan in mediator mass for fixed dark matter mass and mediator couplings; $\MDM=600$ GeV, $\GQ=0.375$ and $\GDM = 1$.}
\label{fig:ATLASwjet}
\end{figure}

Moving on to the `electroweak' final states discussed in Section~\ref{sec:measurements}, Figure~\ref{fig:ATLASwjet} illustrates the sensitivity of vector-boson-plus-jet
($V+$jet) measurements to this model, in this case the dijet mass differential cross section in $W+$-jet events. Strictly speaking, the measurement is made for
events with a single charged lepton, \MET, and jets, interpreted as $W$+jets in the SM. In the BSM model considered here, \MET could in principle
also arise from the dark matter candidate. However, inspections shows that the sensitivity, which is at mediator masses below around a TeV or so, arises
from genuine $W$ bosons produced in association with the mediator --- see Fig.~\ref{fig:plusV} --- which is not a signature typically considered in
constraints on this class of model. The sensitivity
is obviously highly dependent upon the bin width chosen in the SM measurement, which is driven mainly by the dijet mass resolution, although at high masses also by
the number of events in the data.

\begin{figure}
\centering
      \includegraphics[width=0.9\textwidth]{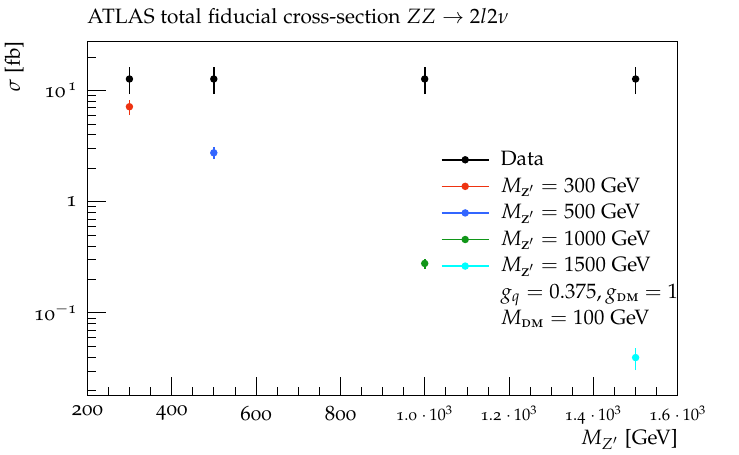}
\caption{Outputs from \rivet for a potential measurement to be included in the limit setting process. Simulated signals for a
sample of mediator masses, interpreted as perturbations to the $ZZ\rightarrow l^+l^- \MET$ cross section
corresponding to the data as measured by ATLAS at 7 TeV~\cite{Aad:2012awa}. The signals form a 1D parameter space scan in
mediator mass \MZP for fixed dark matter mass and mediator couplings; $\MDM=100$ GeV, $g_{q}=0.25$ and $g_{\textsc{dm}} = 1$.}
\label{fig:ATLASzz}
\end{figure}

Also in the `electroweak' category are the diboson measurements. Here the most sensitive is the ATLAS $ZZ$ measurement, in particular the 7~TeV result, which includes
a fiducial cross section measurement of $pp \rightarrow l^+l^- + \MET$, interpreted in the paper as $pp \rightarrow ZZ \rightarrow l^+l^- \nu \bar{\nu}$, but performed in a
sufficiently model-independent fashion that it has the same sensitivity to the $l^+l^- +$ dark matter channel. This is illustrated in Fig.~\ref{fig:ATLASzz}.
The production diagrams are the same as the $V$-jets case, Fig.\ref{fig:plusV}, but in this case the
mediator decays to dark matter rather than back to quarks.
In the absence of any particle-level measurements in the `missing energy plus jets' category of
Section~\ref{sec:measurements}, this measurement has the best sensitivity to dark matter production for this model. Obviously, measurements at 8~TeV and 13~TeV of this
final state, and indeed of jets+\MET, can be expected to improve the sensitivity significantly.

Finally, although they were scanned in the limit-setting process, the currently available isolated photon measurements do not contribute signficantly to the exclusion
limits for this model.

\section{Limits}\label{sec:limits}

\begin{figure}[hp]
\centering
\hfill
\includegraphics[width=.95\textwidth]{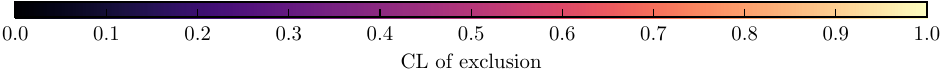}
\centering
	\subfloat[$\GQ=0.25$ and $\GDM = 1$\label{fig:map025}]{%
    \includegraphics[width=0.5\textwidth]{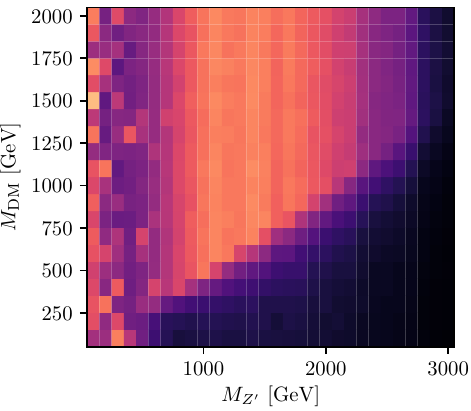}
    }
    \subfloat[$\GQ=0.5$ and $\GDM = 1$\label{fig:map05}]{%
    \includegraphics[width=0.5\textwidth]{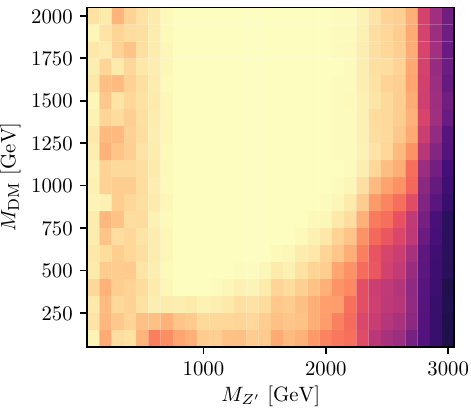}
    }

    \subfloat[$\GQ=0.375$ and $\GDM = 1$\label{fig:map0375}]{%
    \includegraphics[width=0.5\textwidth]{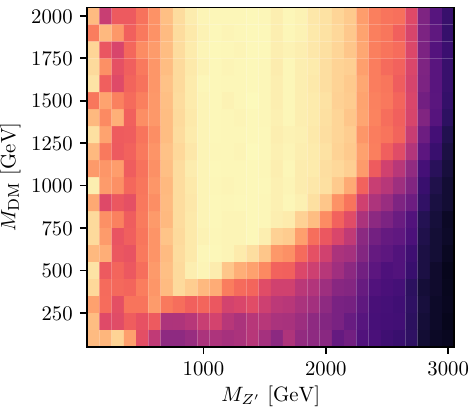}
    }
    \subfloat[$\GQ=0.375$ and $\GDM = 0.25$\label{fig:map0375_025}]{%
    \includegraphics[width=0.5\textwidth]{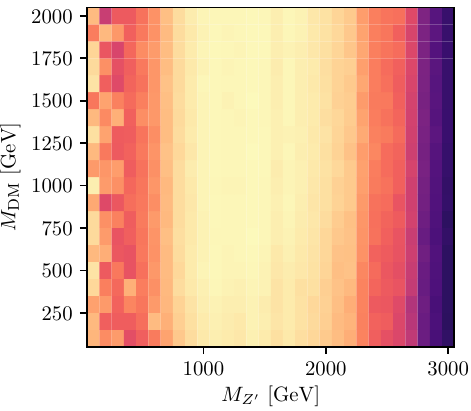}
    }

    \caption{Heatmaps displaying 2D parameter space scans in fixed mass planes corresponding to a fixed $g_{\textsc{dm}} = 1$ and variable $g_{q}$, with
    figure~\ref{fig:map025} representing $g_{q}=0.25$ and figure~\ref{fig:map05} representing $g_{q}=0.5$. The confidence level of exclusion represented
    corresponds to testing the full signal strength hypothesis against the background-only hypothesis, calculated as outlined in
    section~\ref{sec:statmethod}. The combination of measurements entering into the confidence level presented here is the maximally
    sensitive allowed grouping as outlined in section~\ref{sec:dynselec}, considering all available measurements as listed in section~\ref{sec:measurements}.
    (a) Challenging scenario, (b) Optimistic (c) Intermediate (d) DM suppressed.}

    \label{fig:maps}
\end{figure}

The sensitivities derived from multiple distributions such as those discussed in the previous section are combined into `heatmaps' which delineate exclusion
regions and contours in the parameter space of \MDM and \MZP. These are shown in Fig.~\ref{fig:maps} for the four \GQ and \GDM combinations considered.

As expected, the exclusion is much weaker in the `challenging' case and quite strong in the `optimistic' scenario.
For the first three scenarios, at $\MZP > 2\MDM$ the decay of the mediator to dark matter dominates over the decay to jets. This leads to the diagonal structure across the plots,
with the sensitivity above the diagonal, in the left portion of the map, coming mainly from the jet measurements. In the fourth scenario, even when the decay to
DM is kinematically allowed, the jet signatures continue to contribute, and so the diagonal structure is less visible.

At low values of \MZP the sensitivity comes mainly from the $V+$jets signatures. In the challenging scenario, a dip in sensitivity around
$\MZP \approx 700$~GeV is visible, where the sensitivity from inclusive jets and $V+$jets do not quite overlap. In the optimistic scenario, they overlap, and the whole upper
left region of the map is excluded. In addition, the cross section $\times$ branching ratio for quarks $\rightarrow Z^\prime \rightarrow$ quarks remains large
enough that the diagonal cutoff in sensitivity of the jet channels at $\MZP \approx 2\MDM$ is blurred.

To the bottom right region of the diagonal the decay of the mediator to dark matter is kinematically allowed, and for $\GDM=1$ it will dominate
over the decay to quarks. Hence the sensitivity in the inclusive jet (and $V+$jet) signatures drops in all scenarios except the fourth.
This is the region where a measurement of \MET{+}\,jets would be useful
(and indeed it is where the searches performed using such signatures contribute, see, for example, \cite{Kahlhoefer:2015bea,Heisig:2015ira}). Current sensitivity in the intermediate and challenging scenarios
comes from the $l^+l^- + \MET$ measurement, and dies away at $\MZP \approx 750$~GeV. In the fourth scenario, the decays to dark matter are relatively suppressed and
so the $l^+l^- + \MET$ signature makes little contribution. However, as already discussed, the exclusion from the jet measurements remains strong.

\begin{figure}[hp]
\centering
    \subfloat[$\GQ=0.25$ and $\GDM = 1$ \label{fig:cont025}]{%
    \includegraphics[width=0.5\textwidth]{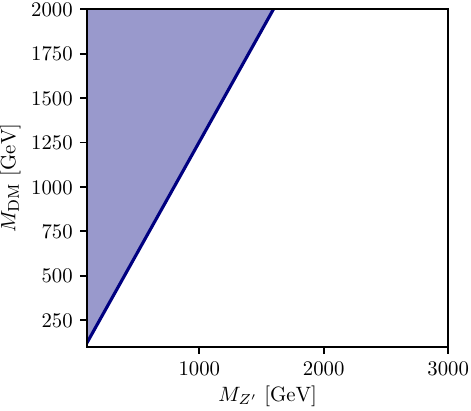}
    }
    \subfloat[$\GQ=0.5$ and $\GDM = 1$\label{fig:cont05}]{%
      \includegraphics[width=0.5\textwidth]{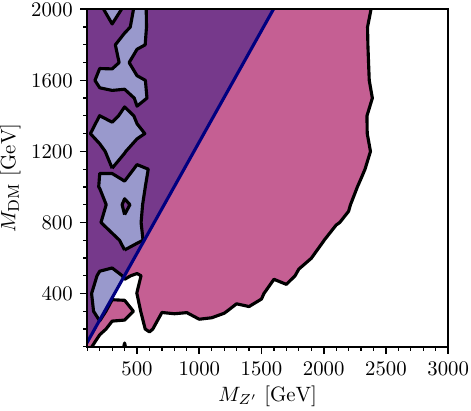}
    }

    \subfloat[$\GQ=0.375$ and $\GDM = 1$\label{fig:cont0375}]{
      \includegraphics[width=0.5\textwidth]{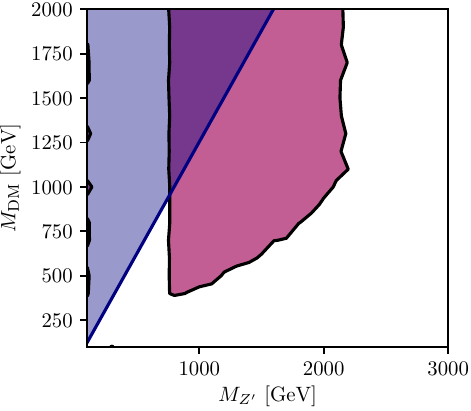}
    }
    \subfloat[$\GQ=0.375$ and $\GDM = 0.25$\label{fig:cont0375_025}]{%
      \includegraphics[width=0.5\textwidth]{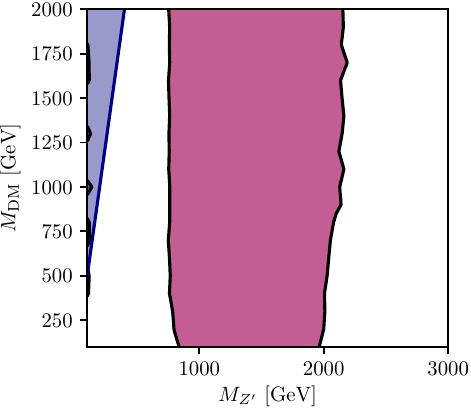}
    }
\caption{Contours in the \MZP and \MDM plane for the considered values of \GDM and \GQ, indicating the excluded region at 90\% confidence level. The triangular shaded
area is the region in which perturbative unitary is violated by the model.}\label{fig:cont}
\end{figure}

The 90\% contours derived from the heatmaps of Fig.~\ref{fig:maps} are shown in Fig.~\ref{fig:cont}.
Note that as expected, the sensitivity from the 7~TeV dijet measurements used here is qualitatively similar, but inferior, to the exclusions obtained
combining the searches in 8~TeV and 13~TeV jet data --- see, for example, \cite{Fairbairn:2016iuf}. This should change once measurements
are available from these later running periods (indeed, the CMS measurement is already made~\cite{Khachatryan:2016wdh}, but is not yet available in \rivet or HepData).
The other channels extend the sensitivity, and this will also improve as more measurements are incorporated.

As mentioned in section~\ref{sec:model},  the parameters of the simplified model are constrained by perturbative unitarity.
In the region $\MDM \gtrsim \sqrt{\pi/2}\, \MZP/g_{\rm DM}$, indicated by the blue shaded area in Fig.~\ref{fig:cont},
the dark matter relic density cannot be calculated reliably~\cite{Kahlhoefer:2015bea}. Since we only consider couplings \GDM and \GQ well within the perturbative regime, perturbative unitarity is respected in the production of mediators at the LHC
and does not provide any further restrictions on the parameter space of our model~\cite{Englert:2016joy}. The physics of dark matter is, of course, constrained by astrophysical and cosmological observations, including in particular the dark matter relic density, and direct and indirect searches for dark matter, see, for example, Refs.~\cite{Kahlhoefer:2015bea, Heisig:2015ira,Jacques:2016dqz} for combined analyses of collider and astrophysical constraints of simplified dark matter models with vector mediators. However, all those constraints are based on additional assumptions on the thermal history of the Universe and astrophysical properties of dark matter, and they do not affect BSM searches at the LHC. Since we have adopted the simplified dark matter model to illustrate the power of the \Contur approach for BSM searches at the LHC in general, rather than providing a detailed cosmological and astrophysical analysis of dark matter, we do not show the corresponding constraints in Fig.~\ref{fig:cont}.

\section{Conclusions}\label{sec:conclusions}

Using a simplified model for weakly-interacting dark matter coupled to the Standard Model via a heavy mediator vector boson, we have developed and demonstrated
a method to efficiently scan existing particle-level measurements from the LHC, implemented in \rivet, to derive expected limits on new physics.
The \Contur method uses measurements which have already been shown to be in good agreement with the SM, and thus is purely aimed at limiting the possibilities
for models of new physics and hopefully narrowing the focus of experimental and theoretical effort on to the best models. It is thus complementary to
direct and dedicated searches. The expected exclusion limits obtained are competitive with limits from searches to  date which have reported null results.
One notable feature is the simultaneous coverage of a wide variety of final states. This leads to enhanced stability of the sensitivity as a function of model
parameters, and also can uncover sensitivity in channels which might not otherwise be considered. For example, in our case unexpected sensitivity is seen in
$V$+jets measurements, as well as the more commonly used dijet and \MET channels. Future plans include better treatment of correlated uncertainties and the incorporation of
SM predictions and uncertainties directly into Contur, rather than relying on previous comparisons.
The method is highly scaleable to new measurements as they are produced, and to new simplified models as they are developed.

\section*{Acknowledgments}
This work started at the `Interdisciplinary Workshop on ‘Models, simulations and data at LHC' in Edinburgh, and continued in the 2015 Les Houches meeting on
TeV-scale physics and two MCnet schools in G\"ottingen. The authors thank the organisers, especially Michela Massimi, Fawzi Boudjema and Steffen Schumann.
They also thank Josh McFayden 
for useful discussions, and STFC for financial support. This work was supported in part by the European Union as part of the FP7 Marie Curie Initial Training Network MCnetITN (PITN-GA-2012-315877). MK is supported by the German Research Foundation DFG through the research unit 2239 ``New physics at the LHC".

\bibliographystyle{JHEP}
\bibliography{simple}

\end{document}